\documentclass[aps,prl,showpacs,reprint,superscriptaddress]{revtex4-1}

\usepackage{graphicx}
\usepackage{dcolumn}
\usepackage{bm}
\usepackage{color}
\usepackage[usenames,dvipsnames,svgnames,table]{xcolor}
\graphicspath{ {Figures/} }
\usepackage{comment}
\usepackage{braket}
\usepackage[]{SIunits}
\usepackage{mathtools}
\usepackage{enumitem}
\usepackage{framed}
\usepackage[version=3]{mhchem}
\usepackage{hyperref}
\PassOptionsToPackage{bookmarks=false}{hyperref}
\usepackage{natbib}

\begin{document}

\title{Evidence of nanoscale Anderson localization induced by intrinsic compositional disorder in InGaN/GaN quantum wells by scanning tunneling luminescence spectroscopy}

\author{W.~Hahn}
\affiliation{Laboratoire de Physique de la Mati\`ere Condens\'ee, Ecole polytechnique, CNRS, Universit\'e Paris Saclay, 91128 Palaiseau Cedex, France}
\affiliation{Experimentelle Physik 2, Technische Universit\"at Dortmund, 44221 Dortmund, Germany}

\author{J.-M.~Lentali}
\affiliation{Laboratoire de Physique de la Mati\`ere Condens\'ee, Ecole polytechnique, CNRS, Universit\'e Paris Saclay, 91128 Palaiseau Cedex, France}

\author{P.~Polovodov}
\affiliation{Laboratoire de Physique de la Mati\`ere Condens\'ee, Ecole polytechnique, CNRS, Universit\'e Paris Saclay, 91128 Palaiseau Cedex, France}

\author{N.~Young}
\affiliation{Materials Department, University of California, Santa Barbara, California 93106, USA}

\author{S.~Nakamura}
\affiliation{Materials Department, University of California, Santa Barbara, California 93106, USA}

\author{J.~S.~Speck}
\affiliation{Materials Department, University of California, Santa Barbara, California 93106, USA}

\author{C.~Weisbuch}
\affiliation{Laboratoire de Physique de la Mati\`ere Condens\'ee, Ecole polytechnique, CNRS, Universit\'e Paris Saclay, 91128 Palaiseau Cedex, France}
\affiliation{Materials Department, University of California, Santa Barbara, California 93106, USA}

\author{M.~Filoche}
\affiliation{Laboratoire de Physique de la Mati\`ere Condens\'ee, Ecole polytechnique, CNRS, Universit\'e Paris Saclay, 91128 Palaiseau Cedex, France}

\author{Y-R.~Wu}
\affiliation{Graduate Institute of Photonics and Optoelectronics and Department of Electrical Engineering, National Taiwan University, Taipei 10617, Taiwan}

\author{M.~Piccardo}
\affiliation{Laboratoire de Physique de la Mati\`ere Condens\'ee, Ecole polytechnique, CNRS, Universit\'e Paris Saclay, 91128 Palaiseau Cedex, France}

\author{F.~Maroun}
\affiliation{Laboratoire de Physique de la Mati\`ere Condens\'ee, Ecole polytechnique, CNRS, Universit\'e Paris Saclay, 91128 Palaiseau Cedex, France}

\author{L.~Martinelli}
\affiliation{Laboratoire de Physique de la Mati\`ere Condens\'ee, Ecole polytechnique, CNRS, Universit\'e Paris Saclay, 91128 Palaiseau Cedex, France}

\author{Y.~Lassailly}
\affiliation{Laboratoire de Physique de la Mati\`ere Condens\'ee, Ecole polytechnique, CNRS, Universit\'e Paris Saclay, 91128 Palaiseau Cedex, France}

\author{J.~Peretti}
\email[]{jacques.peretti@polytechnique.edu}
\affiliation{Laboratoire de Physique de la Mati\`ere Condens\'ee, Ecole polytechnique, CNRS, Universit\'e Paris Saclay, 91128 Palaiseau Cedex, France}

\date{\today}

\begin{abstract}
We present direct experimental evidences of Anderson localization induced by the intrinsic alloy compositional disorder of InGaN/GaN quantum wells. Our approach relies on the measurement of the luminescence spectrum under local injection of electrons from a scanning tunneling microscope tip into a near-surface single quantum well. Fluctuations in the emission line shape are observed on a few-nanometer scale. Narrow emission peaks characteristic of single localized states are resolved. Calculations in the framework of the localization landscape theory provide the effective confining potential map stemming from composition fluctuations. This theory explains well the observed nanometer scale carrier localization and the energies of these Anderson-type localized states. The energy spreading of the emission from localized states is consistent with the usually observed very broad photo- or electro-luminescence spectra of InGaN/GaN quantum well structures.
\end{abstract}

\maketitle

InGaN quantum wells (QWs), the active regions of nitride-based LEDs, display broad photo- or electro-luminescence spectra. The emission is assumed to be inhomogeneously broadened due to contributions from material regions with different eigenenergies associated with fluctuations of the In content. Spatial variations of emission intensities and/or energies were previously observed in nitride semiconductor structures by far field \cite{Okamoto2004} and near-field \cite{Kawakami2016, Ivanov2017} photoluminescence microscopy or cathodoluminescence microscopy \cite{Chichibu1997, Sonderegger2006, Pozina2015}. These observations have evidenced so far emitting domains of either micron or submicron (typically \unit{100}{\nano\meter}) size. While significant energy shifts due to these rather large-scale fluctuations were observed, no linewidths below \unit{100}{meV} at room temperature were detected. These large-scale luminescence fluctuations are usually associated with growth inhomogeneities seen for instance in AFM. However, in the case of random atom positioning in the alloy crystal \cite{Wu2012}, modeling predicts that the compositional disorder should induce Anderson localization in regions of a few {\nano\meter} size \cite{Watson-Parris2011, Yang2014, Schulz2015, Maur2016, Filoche2017, Piccardo2017, Li2017}. Moreover, emission from such single localized states is expected to exhibit a narrow linewidth of a few tens of meV at room temperature, i.e. comparable to what was observed from single GaN quantum dots \cite{Holmes2014}. Such emission features exhibiting size and spectroscopic signatures characteristic of carrier localization induced by the alloy disorder have not been evidenced up to now.

Here we report on the observation of fluctuations in the luminescence spectrum on the scale of a few {\nano\meter} as expected for such localization effects. Our approach is based on scanning tunneling luminescence (STL) microscopy \cite{Samuelson1992} which allows luminescence measurements under local carrier injection with a {\nano\meter} scale resolution. In order to reach the required spatial resolution, we used p-type single InGaN/GaN QW structures with a thin top GaN layer separating the QW from the injecting tip of the scanning tunneling microscope (STM) and we analyzed the spatial variations in the luminescence spectrum when scanning the tip by steps of a few {\nano\meter}. At each tip position a small material volume is excited which contains only a few localization regions. Large changes in the emission line shape are observed over distances as short as the scanning step. Narrow peaks corresponding to the emission from single localized states at room temperature are resolved. Calculations based on the localization landscape theory \cite{Filoche2017, Piccardo2017, Li2017} support the interpretation in terms of compositional disorder-induced localization.
\begin{figure}
\includegraphics[width=0.85\linewidth]{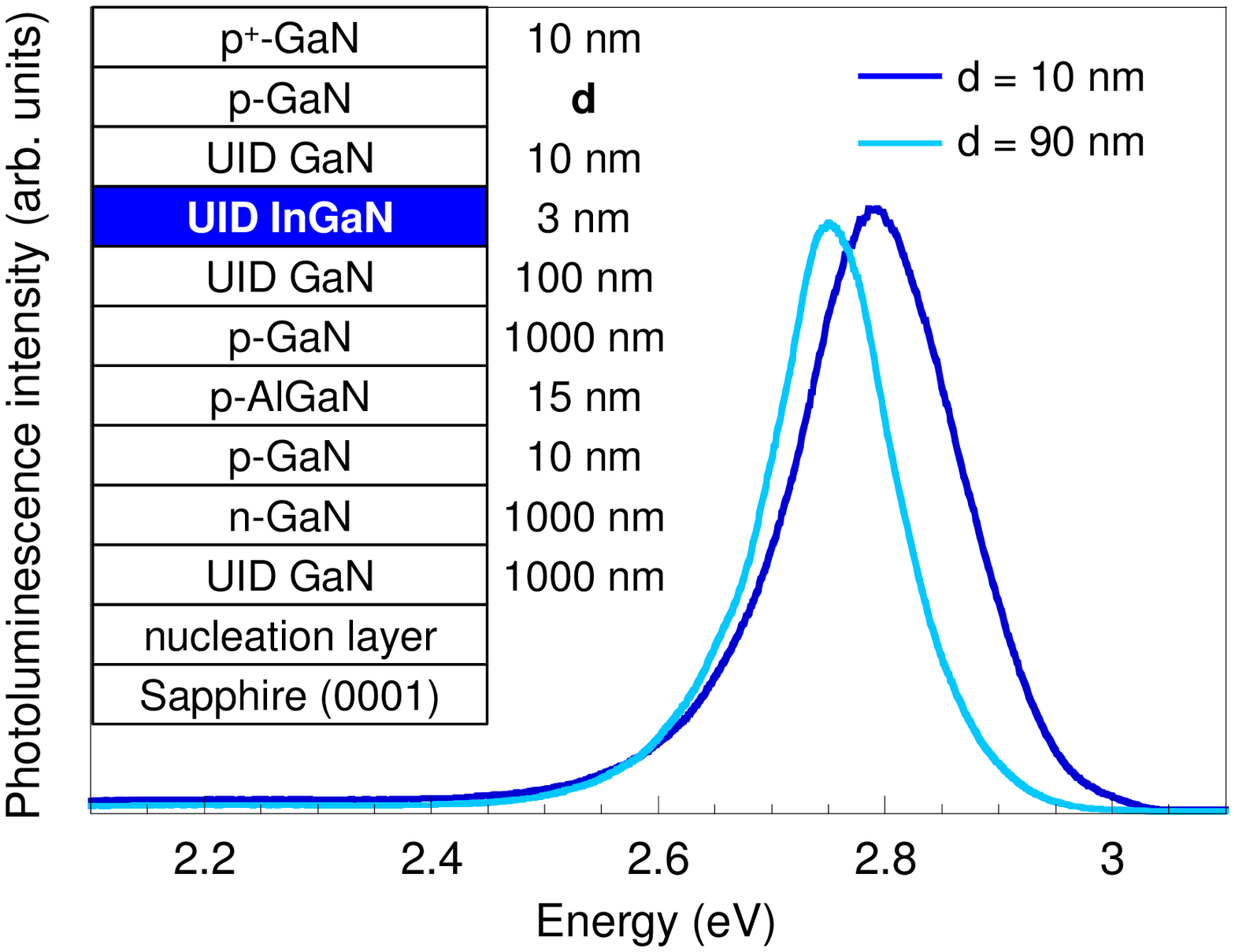}
\caption{Photoluminescence spectra of InGaN/GaN QW samples at \unit{300}{\kelvin} for a \unit{394}{{\nano\meter}} excitation wavelength. Inset: sample structure; the thickness $d$ of the p-GaN layer, below the p$^{+}$-GaN cap, is \unit{10}{{\nano\meter}} (resp. \unit{90}{{\nano\meter}}) in S10 (resp. S90).}
\label{fig:PL}
\end{figure}
The samples schematized in Fig.~\ref{fig:PL} (inset) were grown by metalorganic chemical vapor deposition on a \unit{0.2}{\degree} miscut (0001) sapphire substrate. The InGaN QW is separated from the surface by a GaN three-layer stack consisting of an unintentionally doped (UID) spacer, a p-doped layer and a p$^{+}$-overdoped cap. Two structures, hereafter referred to as S10 and S90, were grown, respectively, with a \unit{10}{\nano\meter} and \unit{90}{\nano\meter}-thick p-GaN layer. The nominal In content in the \unit{3}{\nano\meter} thick InGaN QW is \unit{18}{\%}. The photoluminescence spectra of the two samples (Fig.~\ref{fig:PL}) exhibit the characteristic blue emission with a slight shift due to different actual sample compositions and/or to large scale inhomogeneities. The full width at half maximum (FWHM) of these two spectra is of about \unit{150}{meV}, as usually observed. A patch ohmic contact was processed on the GaN surface in order to apply the tunneling bias $V_b$ between the STM tip and the sample. After chemical cleaning and passivation \cite{Tereshchenko2004}, the STM images of the sample surfaces (see Supplemental Material) exhibited $\sim$\unit{150}{\nano\meter}-wide atomically flat terraces separated by (bi-)atomic steps and hexagonal pits (in density of \unit{\sim 10^9}{cm^{-2}}) characteristic of emerging dislocations on c-plane GaN \cite{Visconti2000}. Pits-free areas were selected for STL measurements to avoid artefacts originating from such defects.
STL experiments were performed at room temperature in the constant tunneling current mode. Electrons were locally injected into the grounded sample from the negatively biased STM tip. A \unit{0.6} aperture lens and a \unit{0.6}{\milli\meter} core diameter optical fiber were used to collect and guide the luminescence light from the sample backside to the spectrometer. With a fully opened entrance slit the spectrometer resolution was \unit{35}{meV} FWHM.
\begin{figure}
\includegraphics[width=0.85\linewidth]{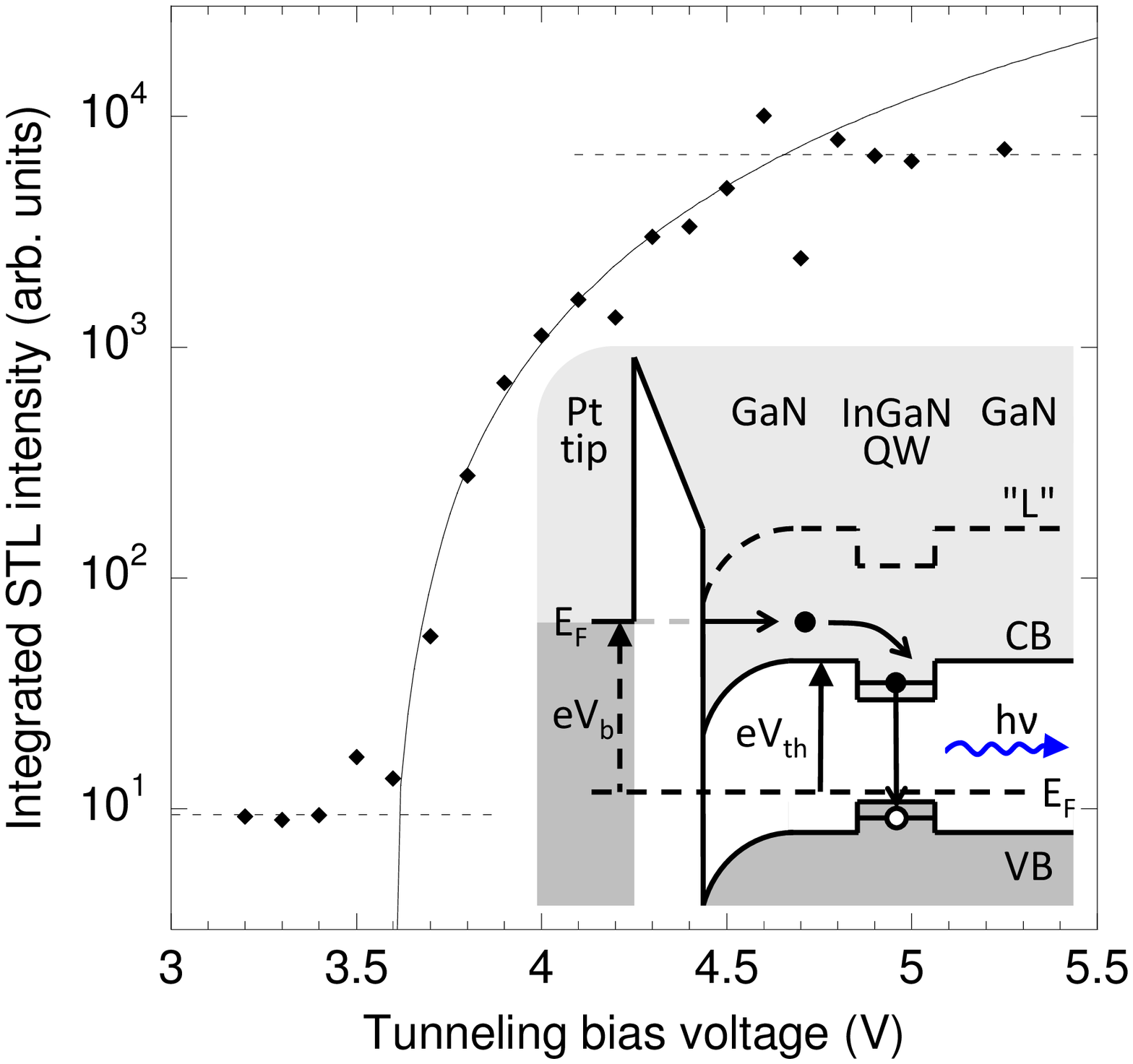}
\caption{Spectrally integrated luminescence intensity as a function of $V_b$ for a constant tunneling current of \unit{20}{\nano\ampere}. The line is a fit by a function of $(V_b - V_{th})^2$ with $V_{th}=$ \unit{3.6}{\volt}. Inset: schematics of the tip-sample band diagram in real space; the extrema of the semiconductor valence band (VB), conduction band (CB) and conduction side valley (L) are represented.}
\label{fig:I-V}
\end{figure}
The variation of the spectrally integrated QW luminescence signal as a function of $V_b$ is plotted in Fig.~\ref{fig:I-V}. Similar variations were obtained on both S10 and S90 (no light was detected in the QW emission range from a control p-GaN sample without a QW). For sub-band-gap injection, no luminescence was detected. The luminescence onset threshold, $V_{th} \approx$ \unit{3.6}{\volt}, almost corresponds to the \unit{3.25}{e\volt} energy separation between the Fermi level and the conduction band minimum in the p-GaN layer (see inset in Fig.~\ref{fig:I-V}). This slight difference is most probably due to a potential drop in the contact/access resistance. Beyond $V_{th}$, the luminescence intensity steadily increases up to \unit{4.5}{\volt}, following the usual quadratic variation in $(V_b - V_{th})^2$ \cite{Renaud1991}. Beyond \unit{4.5}{\volt} (i.e. $\sim$\unit{1}{\volt} above $V_{th}$) a plateau is observed which may result from the electron injection and transport into upper conduction valleys in the bulk region \cite{Piccardo2014, Marcinkevicius2016}. The saturation of the luminescence signal might be due to a poorer capture of electrons from these valleys into the QW or to the different angle of refraction of electrons injected in the side valleys \cite{Prietsch1995}.

\begin{figure}
\includegraphics[width=0.85\linewidth]{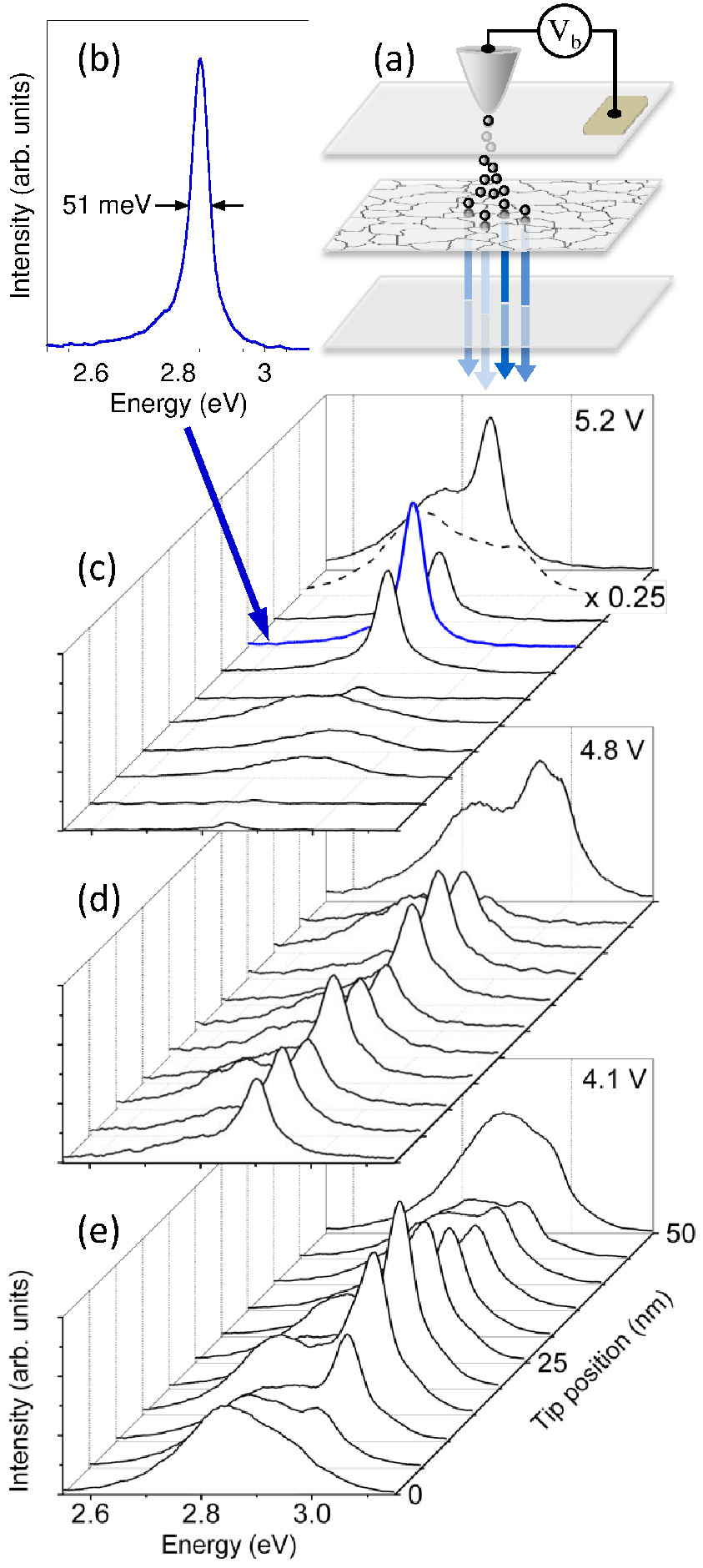}
\caption{(a) Schematic of the STL experiment showing the electron injection from the STM tip into the QW which exhibits localization regions. (b) Narrow STL spectrum characteristic of the emission from a single localized state. (c-e) Evolution of the STL spectrum along \unit{50}{\nano\meter} line scans measured at different locations on S10 with constant injection current and bias of \unit{2}{\nano\ampere} and (c) \unit{4.1}{\volt}, (d) \unit{4.8}{\volt}, (e) \unit{5.2}{\volt}.}    
\label{fig:zoom}
\end{figure}

The spectrally integrated luminescence intensity maps (not shown here) recorded when scanning the tip over the surface exhibit fluctuations over more than one order of magnitude, with very localized features. However, spatial inhomogeneities in the luminescence intensity map are not specific for localization effects in the QW since they may have different other microscopic origin (injection efficiency, transport processes, non-radiative recombination, interface transmission...). In contrast, the emission spectrum line shape is specifically related to the recombination process and carry direct signature of the local potential fluctuations seen by the electrons in the QW (Fig. ~\ref{fig:zoom}(a)). We therefore focus on the analysis of the spatial fluctuations of the luminescence spectrum rather than of the luminescence intensity.
Fig.~\ref{fig:zoom}(c-e) shows three characteristic examples of the evolution of the locally excited STL spectrum emitted from S10 when scanning the tip by steps of \unit{5}{\nano\meter}. At each tip position the emission spectrum was recorded. These three experiments were performed with the same tunneling current of \unit{2}{\nano\ampere} but at three different tunneling biases and locations on the sample. In all three experiments, the spectra show significant fluctuations in intensity, energy and shape with multiple peaks rising up and vanishing along the line scan. Remarkably, narrow emission lines are resolved which exhibit FWHM of $\sim$\unit{50}{meV} (Fig.~\ref{fig:zoom}(b)), similar to the \unit{36}{meV} linewidth measured on single GaN quantum dots at room temperature \cite{Holmes2014} when taking into account the \unit{35}{meV} resolution of our optical spectroscopy set-up. These narrow contributions are therefore fully compatible with the expected emission from a single localized state. The relative intensities of the narrow contributions vary significantly over distances as short as the \unit{5}{\nano\meter} scanning step. The most intense narrow contribution in Fig.~\ref{fig:zoom}(c) attenuates over a distance of about \unit{10}{\nano\meter}. This indicates that the characteristic size of the carrier localization domains is of a few {\nano\meter} and that electrons travel at most over a few localized states. The observed short in-plane transport distance suggests that the carrier lifetime in the QW is of the order of the transfer time between localized states. The transfer time between neighboring localized states by phonon-assisted tunneling can be of hundreds of {\pico\second} (as was measured in GaAs coupled QWs \cite{Oberli1990, Tada1988}), while saturated electron lifetime corresponding to occupied states of the same order of magnitude (\unit{300}{\pico\second}) was measured in InGaN QWs \cite{Casey1976,David2010}. This somehow contrasts with sizeable diffusion in InGaN QWs deduced from photoluminescence microscopy techniques \citep{Danhof2011, Mensi2018} which probed transport phenomena over much larger scales corresponding to carriers in delocalized states. However, the coexistence of non-diffusing and diffusing carriers was already observed in GaAs QWs \citep{Hegarty1985}.

The fact that narrow emission contributions corresponding to single localized states can be detected shows that the STL spatial resolution on S10 is comparable with the size of the localization regions. Injected electrons first experience the band bending region (BBR) and it was demonstrated \cite{Renaud1991} that, similarly to the ballistic electron emission microscopy configuration \cite{Kaiser2001}, overcoming the BBR barrier selects the transmission of ballistic electrons with small transverse wavevector and allows STL resolution at the {\nano\meter} level. Beyond the BBR, a short remaining distance to the QW limits defocusing. In addition, forward-transported electrons experience the shortest path and then have a higher probability to reach the QW. Note that a STL geometry similar to ours was used for studying near-surface InGaAs QWs in InP \cite{Samuelson1992}. Fluctuations in the STL spectrum on a scale of a few tens of {\nano\meter} were observed but resolution was most probably not limited by electron spreading but rather by the size of the InGaAs/GaAs QW interface islands under observation which are known to extend over tens of {\nano\meter}.

\begin{figure}
\includegraphics[width=\linewidth]{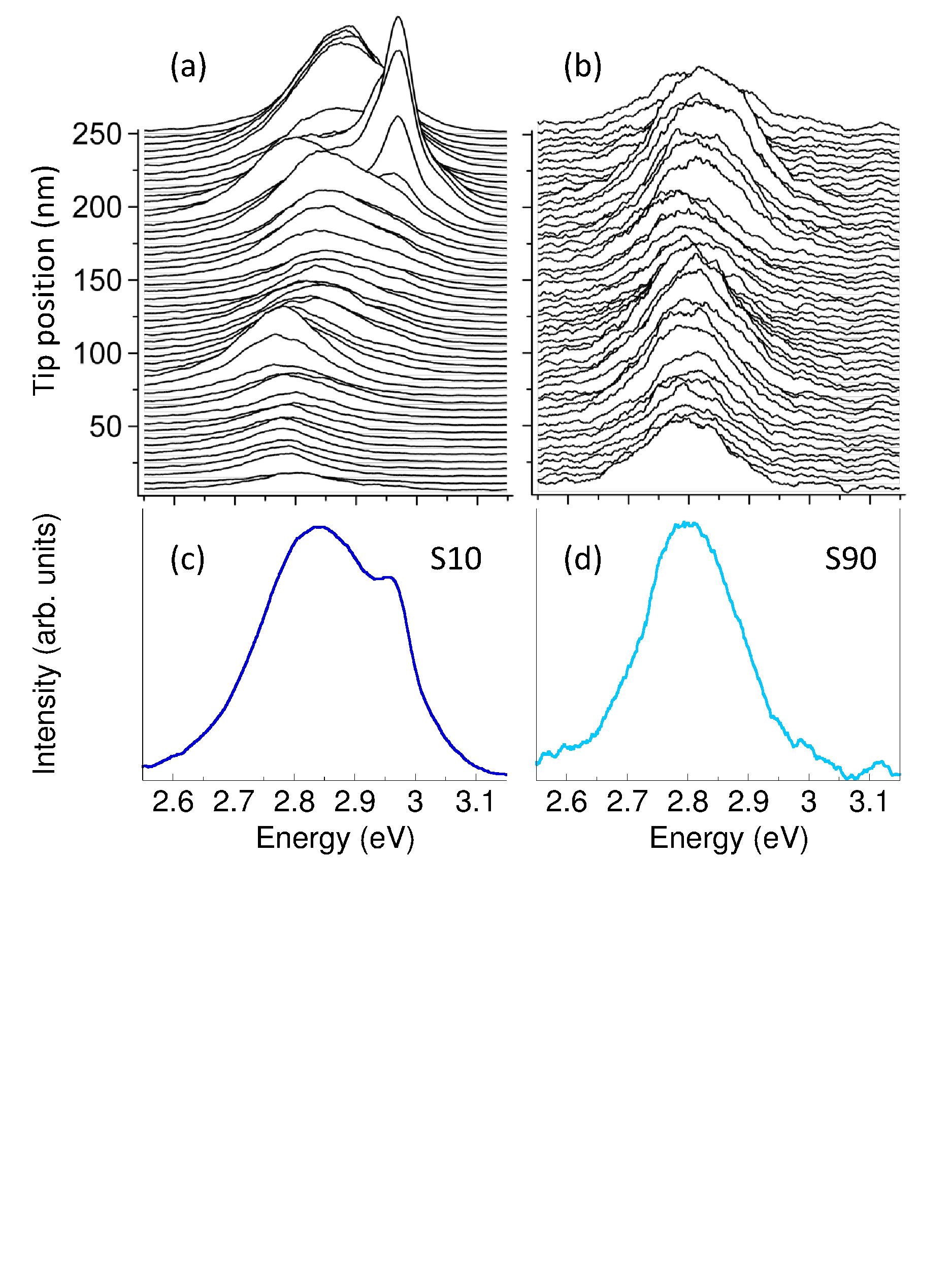}
\caption{Evolution of the STL spectrum measured on (a) S10 and (b) S90 along a \unit{250}{\nano\meter} line scan by \unit{5}{\nano\meter} steps for constant tunneling current and bias of \unit{2}{\nano\ampere} and \unit{4.1}{\volt}. (c-d) Sum of the \unit{50} local spectra for S10 and S90, respectively.}
\label{fig:linescan}
\end{figure}

When the remaining distance to cross from the BBR to the QW is large, diffusion should deteriorate the resolution. Significant differences between S10 and S90 are thus expected. Figure~\ref{fig:linescan} compares the evolution of the STL spectrum along a \unit{250}{\nano\meter} line scan performed on S10 and S90 for constant injection current and bias of \unit{2}{\nano\ampere} and \unit{4.1}{\volt}, respectively.
On S10 (Fig.~\ref{fig:linescan}(a)), as already discussed, the measured spectra evidence significant fluctuations with multiple narrow contributions showing up in different injected area. The sum of the \unit{50} spectra (Fig.~\ref{fig:linescan}(c)) exhibits a line shape that tends to the \unit{150}{meV}-wide spatially integrated photoluminescence spectrum (Fig.~\ref{fig:PL}). This indicates that the emission from an area of the order of \unit{1000}{\nano\square\meter} (assuming a spatial resolution of about \unit{5}{\nano\meter} over the \unit{250}{\nano\meter} line scan) already averages over many localized states. Thus, the previous observations of rather large-scale fluctuations and broad linewidths that were obtained with lower spatial resolution \cite{Chichibu1997, Okamoto2004, Sonderegger2006, Pozina2015, Kawakami2016, Ivanov2017} are indeed probably related to sample growth inhomogeneities (such as may be associated with fluctuations in the step spacing on the growth surface) and not to carrier localization induced by the intrinsic alloy disorder. Note that, the large-scale growth-induced fluctuations can be at the origin of the usual asymmetric line shape of the emission of InGaN/GaN structures (as seen in \cite{Okamoto2004} from the submicron analysis of green emitting structures).
The spectra measured along a line scan on S90 (Fig.~\ref{fig:linescan}(b)) do not show narrow emission lines. A single broad peak is systematically observed which exhibits much less spatial fluctuations than on S10. This is consistent with the expected electron spreading during the diffusive transport from the BBR to the QW so that, many localization regions are excited at once in the QW for a given tip position. Emission fluctuations are thus already largely averaged in a single locally excited spectrum and changing the tip position by a small step only slightly affects the spectrum. As a consequence, all the spectra measured along the line scan almost coincide and are very similar (in position and width) to the summed spectrum (Fig.~\ref{fig:linescan}(d)). Note that, previous STL microscopy studies in nitride devices showed luminescence intensity fluctuations over a typical scale of \unit{100}{\nano\meter}. In these early experiments, structures were not compatible with high resolution because of thick injection layers (\unit{100}{\nano\meter} in \cite{Evoy1999}, \unit{200}{\nano\meter} in \cite{Manson-Smith2001}) and multiple QW active region \cite{Evoy1999} or a single QW capped with a GaN/AlGaN barrier \cite{Manson-Smith2001}.
\begin{figure}
\includegraphics[width=\linewidth]{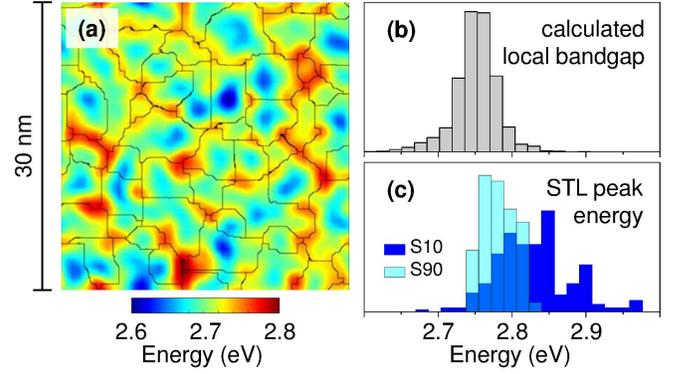}
\caption{(a) Calculated \unit{30}{\nano\meter} $\times$ \unit{30}{\nano\meter} map of the localization function $1/u$ at the QW center plane in a structure corresponding to S10. The black lines indicate the crest lines which bound the localization regions. (b) Histogram of the local band gaps calculated from the overlap of the $1/u$ maps of electrons and holes. (c) Histograms of the STL peak energies measured along six different linescans.}
\label{fig:landscape} 
\end{figure}

The scale of \unit{5}{\nano\meter} (the scanning step) over which significant changes in the STL spectrum are observed on S10 is consistent with the typical size of the localization regions predicted by the localization landscape theory \cite{Filoche2017, Piccardo2017, Li2017}. According to this theory, the local density of quantum states in a random potential can be obtained without having to solve the Schr\"odinger equation $\hat{H}\psi=E\psi$, which is an eigenvalue problem. Instead, one has to solve a much simpler partial differential equation with uniform right hand side, the associated Dirichlet problem: $\hat{H} u=1$ \cite{Filoche2012, Arnold2016}. The solution called the localization landscape function $u(\bf{r})$, is a positive valued function. It has been shown that the localized eigenfunctions are essentially confined within the valley lines of $u$, or equivalently within the crest lines of $1/u$ which acts as an effective potential. Figure~\ref{fig:landscape}(a) shows a 2D cut of the calculated conduction $1/u$ map at the center plane of the QW in a simulated structure corresponding to S10. The black lines follow the crest of the calculated potential landscape and delimit the electron localization regions which exhibit typical size of \unit{4} to \unit{5}{\nano\meter}. The fundamental energy in each localization region $\Omega_m$ can be estimated \cite{Filoche2017} using the landscape function u only:
\begin{equation}
E_{0}^{(m)} \approx \frac{\iiint_{\Omega_m} u(\bf{r}) \, \mathrm d^3r}{\iiint_{\Omega_m} u^2(\bf{r}) \, \mathrm d^3r}.
\label{eq:element}
\end{equation}
A similar localization landscape is obtained for holes and each region resulting from the overlap of electron and hole landscapes is expected to emit a narrow spectrum with a characteristic transition energy corresponding to the local effective bandgap $E_{0e}^{(m)}+E_{0h}^{(n)}$ for the $m$ and $n$ electron and hole overlapping regions. Fig. 5(b) shows the histogram of the local effective bandgaps calculated for ten sets of random distribution of In atoms in a \unit{30}{\nano\meter} $\times$ \unit{30}{\nano\meter} area of the simulated InGaN/GaN structure. The effective bandgaps spread over more than \unit{200}{meV}. In Fig.~\ref{fig:landscape}(c) are plotted the histograms of the STL peak energy measured along six line scans on S10 and S90. For spectra showing multiple contributions, only the most intense one is considered. The fluctuations of the peak energy for S10 span a range equivalent to the calculated local effective bandgap fluctuations. For S90, the histogram is more sharply peaked around the average emission energy since each locally measured spectrum already averages the emission of many localization regions.

In conclusion, we have observed for the first time by STL micro-spectroscopy the emission from single Anderson-type localized states which form the inhomogeneously broadened InGaN QW emission. The observed variations in the emission spectra over distances as short as \unit{5} {\nano\meter} are consistent with the fluctuations expected to be induced by the random compositional disorder of the QW ternary alloy. Both the spatial and energy scales of these fluctuations are well described by the localization landscape theory. Moreover, the decrease in the emission of single localization regions over typically \unit{10}{{\nano\meter}} indicates that in-plane transfer between localized states operates only over very short distances. This approach can be further applied to assess the impact of disorder as a function of alloying or to probe electronic processes at the nanometer scale in near-surface luminescent structures of other materials exhibiting disorder of different origins. For instance, studying transport in the well-known GaAs/AlGaAs QW system whose disorder originates from the interface steps could allow identifying the mechanism leading to the mobility edge \cite{Hegarty1985}.

We thank Aur\'{e}lien David for fruitful discussions. This work was supported by the French National Research Agency (ANR) and Taiwanese Ministry of Science and Technology (MOST) (CRIPRONI project : Grants No. ANR-14-CE05-0048-01, No. MOST-104-2923-E-002-004-MY3, and No. MOST-105-2221-E-002-098-MY3), and by ARPA-E, U.S. Department of Energy (program DE-EE0007096), the UCSB Solid State Lighting and Energy Electronics Center, and KACST-KAUST-UCSB Solid State Lighting Program (SSLP).

\bibliography{biblioSTL1}

\begin{thebibliography}{34}%
\makeatletter
\providecommand \@ifxundefined [1]{%
 \@ifx{#1\undefined}
}%
\providecommand \@ifnum [1]{%
 \ifnum #1\expandafter \@firstoftwo
 \else \expandafter \@secondoftwo
 \fi
}%
\providecommand \@ifx [1]{%
 \ifx #1\expandafter \@firstoftwo
 \else \expandafter \@secondoftwo
 \fi
}%
\providecommand \natexlab [1]{#1}%
\providecommand \enquote  [1]{``#1''}%
\providecommand \bibnamefont  [1]{#1}%
\providecommand \bibfnamefont [1]{#1}%
\providecommand \citenamefont [1]{#1}%
\providecommand \href@noop [0]{\@secondoftwo}%
\providecommand \href [0]{\begingroup \@sanitize@url \@href}%
\providecommand \@href[1]{\@@startlink{#1}\@@href}%
\providecommand \@@href[1]{\endgroup#1\@@endlink}%
\providecommand \@sanitize@url [0]{\catcode `\\12\catcode `\$12\catcode
  `\&12\catcode `\#12\catcode `\^12\catcode `\_12\catcode `\%12\relax}%
\providecommand \@@startlink[1]{}%
\providecommand \@@endlink[0]{}%
\providecommand \url  [0]{\begingroup\@sanitize@url \@url }%
\providecommand \@url [1]{\endgroup\@href {#1}{\urlprefix }}%
\providecommand \urlprefix  [0]{URL }%
\providecommand \Eprint [0]{\href }%
\providecommand \doibase [0]{http://dx.doi.org/}%
\providecommand \selectlanguage [0]{\@gobble}%
\providecommand \bibinfo  [0]{\@secondoftwo}%
\providecommand \bibfield  [0]{\@secondoftwo}%
\providecommand \translation [1]{[#1]}%
\providecommand \BibitemOpen [0]{}%
\providecommand \bibitemStop [0]{}%
\providecommand \bibitemNoStop [0]{.\EOS\space}%
\providecommand \EOS [0]{\spacefactor3000\relax}%
\providecommand \BibitemShut  [1]{\csname bibitem#1\endcsname}%
\let\auto@bib@innerbib\@empty
\bibitem [{\citenamefont {Okamoto}\ \emph {et~al.}(2004)\citenamefont
  {Okamoto}, \citenamefont {Choi}, \citenamefont {Kawakami}, \citenamefont
  {Terazima}, \citenamefont {Mukai},\ and\ \citenamefont
  {Fujita}}]{Okamoto2004}%
  \BibitemOpen
  \bibfield  {author} {\bibinfo {author} {\bibfnamefont {K.}~\bibnamefont
  {Okamoto}}, \bibinfo {author} {\bibfnamefont {J.}~\bibnamefont {Choi}},
  \bibinfo {author} {\bibfnamefont {Y.}~\bibnamefont {Kawakami}}, \bibinfo
  {author} {\bibfnamefont {M.}~\bibnamefont {Terazima}}, \bibinfo {author}
  {\bibfnamefont {T.}~\bibnamefont {Mukai}}, \ and\ \bibinfo {author}
  {\bibfnamefont {S.}~\bibnamefont {Fujita}},\ }\href {\doibase
  10.1143/JJAP.43.839} {\bibfield  {journal} {\bibinfo  {journal} {Jpn. J.
  Appl. Phys. 1}\ }\textbf {\bibinfo {volume} {43}},\ \bibinfo {pages} {839}
  (\bibinfo {year} {2004})}\BibitemShut {NoStop}%
\bibitem [{\citenamefont {Kawakami}\ \emph {et~al.}(2016)\citenamefont
  {Kawakami}, \citenamefont {Kaneta}, \citenamefont {Hashiya},\ and\
  \citenamefont {Funato}}]{Kawakami2016}%
  \BibitemOpen
  \bibfield  {author} {\bibinfo {author} {\bibfnamefont {Y.}~\bibnamefont
  {Kawakami}}, \bibinfo {author} {\bibfnamefont {A.}~\bibnamefont {Kaneta}},
  \bibinfo {author} {\bibfnamefont {A.}~\bibnamefont {Hashiya}}, \ and\
  \bibinfo {author} {\bibfnamefont {M.}~\bibnamefont {Funato}},\ }\href
  {\doibase 10.1103/PhysRevApplied.6.044018} {\bibfield  {journal} {\bibinfo
  {journal} {Phys. Rev. Appl.}\ }\textbf {\bibinfo {volume} {6}},\ \bibinfo
  {pages} {044018} (\bibinfo {year} {2016})}\BibitemShut {NoStop}%
\bibitem [{\citenamefont {Ivanov}\ \emph {et~al.}(2017)\citenamefont {Ivanov},
  \citenamefont {Marcinkevicius}, \citenamefont {Mensi}, \citenamefont
  {Martinez}, \citenamefont {Kuritzky}, \citenamefont {Myers}, \citenamefont
  {Nakamura},\ and\ \citenamefont {Speck}}]{Ivanov2017}%
  \BibitemOpen
  \bibfield  {author} {\bibinfo {author} {\bibfnamefont {R.}~\bibnamefont
  {Ivanov}}, \bibinfo {author} {\bibfnamefont {S.}~\bibnamefont
  {Marcinkevicius}}, \bibinfo {author} {\bibfnamefont {M.~D.}\ \bibnamefont
  {Mensi}}, \bibinfo {author} {\bibfnamefont {O.}~\bibnamefont {Martinez}},
  \bibinfo {author} {\bibfnamefont {L.~Y.}\ \bibnamefont {Kuritzky}}, \bibinfo
  {author} {\bibfnamefont {D.~J.}\ \bibnamefont {Myers}}, \bibinfo {author}
  {\bibfnamefont {S.}~\bibnamefont {Nakamura}}, \ and\ \bibinfo {author}
  {\bibfnamefont {J.~S.}\ \bibnamefont {Speck}},\ }\href {\doibase
  10.1103/PhysRevApplied.7.064033} {\bibfield  {journal} {\bibinfo  {journal}
  {Phys. Rev. Appl.}\ }\textbf {\bibinfo {volume} {7}},\ \bibinfo {pages}
  {064033} (\bibinfo {year} {2017})}\BibitemShut {NoStop}%
\bibitem [{\citenamefont {Chichibu}\ \emph {et~al.}(1997)\citenamefont
  {Chichibu}, \citenamefont {Wada},\ and\ \citenamefont
  {Nakamura}}]{Chichibu1997}%
  \BibitemOpen
  \bibfield  {author} {\bibinfo {author} {\bibfnamefont {S.}~\bibnamefont
  {Chichibu}}, \bibinfo {author} {\bibfnamefont {K.}~\bibnamefont {Wada}}, \
  and\ \bibinfo {author} {\bibfnamefont {S.}~\bibnamefont {Nakamura}},\ }\href
  {\doibase 10.1063/1.120025} {\bibfield  {journal} {\bibinfo  {journal} {Appl.
  Phys. Lett.}\ }\textbf {\bibinfo {volume} {71}},\ \bibinfo {pages} {2346}
  (\bibinfo {year} {1997})}\BibitemShut {NoStop}%
\bibitem [{\citenamefont {Sonderegger}\ \emph {et~al.}(2006)\citenamefont
  {Sonderegger}, \citenamefont {Feltin}, \citenamefont {Merano}, \citenamefont
  {Crottini}, \citenamefont {Carlin}, \citenamefont {Sachot}, \citenamefont
  {Deveaud}, \citenamefont {Grandjean},\ and\ \citenamefont
  {Ganière}}]{Sonderegger2006}%
  \BibitemOpen
  \bibfield  {author} {\bibinfo {author} {\bibfnamefont {S.}~\bibnamefont
  {Sonderegger}}, \bibinfo {author} {\bibfnamefont {E.}~\bibnamefont {Feltin}},
  \bibinfo {author} {\bibfnamefont {M.}~\bibnamefont {Merano}}, \bibinfo
  {author} {\bibfnamefont {A.}~\bibnamefont {Crottini}}, \bibinfo {author}
  {\bibfnamefont {J.~F.}\ \bibnamefont {Carlin}}, \bibinfo {author}
  {\bibfnamefont {R.}~\bibnamefont {Sachot}}, \bibinfo {author} {\bibfnamefont
  {B.}~\bibnamefont {Deveaud}}, \bibinfo {author} {\bibfnamefont
  {N.}~\bibnamefont {Grandjean}}, \ and\ \bibinfo {author} {\bibfnamefont
  {J.~D.}\ \bibnamefont {Ganière}},\ }\href {\doibase 10.1063/1.2397562}
  {\bibfield  {journal} {\bibinfo  {journal} {Appl. Phys. Lett.}\ }\textbf
  {\bibinfo {volume} {89}},\ \bibinfo {pages} {232109} (\bibinfo {year}
  {2006})}\BibitemShut {NoStop}%
\bibitem [{\citenamefont {Pozina}\ \emph {et~al.}(2015)\citenamefont {Pozina},
  \citenamefont {Ciechonski}, \citenamefont {Bi}, \citenamefont {Samuelson},\
  and\ \citenamefont {Monemar}}]{Pozina2015}%
  \BibitemOpen
  \bibfield  {author} {\bibinfo {author} {\bibfnamefont {G.}~\bibnamefont
  {Pozina}}, \bibinfo {author} {\bibfnamefont {R.}~\bibnamefont {Ciechonski}},
  \bibinfo {author} {\bibfnamefont {Z.}~\bibnamefont {Bi}}, \bibinfo {author}
  {\bibfnamefont {L.}~\bibnamefont {Samuelson}}, \ and\ \bibinfo {author}
  {\bibfnamefont {B.}~\bibnamefont {Monemar}},\ }\href {\doibase
  10.1063/1.4938208} {\bibfield  {journal} {\bibinfo  {journal} {Appl. Phys.
  Lett.}\ }\textbf {\bibinfo {volume} {107}},\ \bibinfo {pages} {251106}
  (\bibinfo {year} {2015})}\BibitemShut {NoStop}%
\bibitem [{\citenamefont {Wu}\ \emph {et~al.}(2012)\citenamefont {Wu},
  \citenamefont {Shivaraman}, \citenamefont {Wang},\ and\ \citenamefont
  {Speck}}]{Wu2012}%
  \BibitemOpen
  \bibfield  {author} {\bibinfo {author} {\bibfnamefont {Y.~R.}\ \bibnamefont
  {Wu}}, \bibinfo {author} {\bibfnamefont {R.}~\bibnamefont {Shivaraman}},
  \bibinfo {author} {\bibfnamefont {K.~C.}\ \bibnamefont {Wang}}, \ and\
  \bibinfo {author} {\bibfnamefont {J.~S.}\ \bibnamefont {Speck}},\ }\href
  {\doibase 10.1063/1.4747532} {\bibfield  {journal} {\bibinfo  {journal}
  {Appl. Phys. Lett.}\ }\textbf {\bibinfo {volume} {101}},\ \bibinfo {pages}
  {083505} (\bibinfo {year} {2012})}\BibitemShut {NoStop}%
\bibitem [{\citenamefont {Watson-Parris}\ \emph {et~al.}(2011)\citenamefont
  {Watson-Parris}, \citenamefont {Godfrey}, \citenamefont {Dawson},
  \citenamefont {Oliver}, \citenamefont {Galtrey}, \citenamefont {Kappers},\
  and\ \citenamefont {Humphreys}}]{Watson-Parris2011}%
  \BibitemOpen
  \bibfield  {author} {\bibinfo {author} {\bibfnamefont {D.}~\bibnamefont
  {Watson-Parris}}, \bibinfo {author} {\bibfnamefont {M.~J.}\ \bibnamefont
  {Godfrey}}, \bibinfo {author} {\bibfnamefont {P.}~\bibnamefont {Dawson}},
  \bibinfo {author} {\bibfnamefont {R.~A.}\ \bibnamefont {Oliver}}, \bibinfo
  {author} {\bibfnamefont {M.~J.}\ \bibnamefont {Galtrey}}, \bibinfo {author}
  {\bibfnamefont {M.~J.}\ \bibnamefont {Kappers}}, \ and\ \bibinfo {author}
  {\bibfnamefont {C.~J.}\ \bibnamefont {Humphreys}},\ }\href {\doibase
  10.1103/PhysRevB.83.115321} {\bibfield  {journal} {\bibinfo  {journal} {Phys.
  Rev. B}\ }\textbf {\bibinfo {volume} {83}},\ \bibinfo {pages} {115321}
  (\bibinfo {year} {2011})}\BibitemShut {NoStop}%
\bibitem [{\citenamefont {Yang}\ \emph {et~al.}(2014)\citenamefont {Yang},
  \citenamefont {Shivaraman}, \citenamefont {Speck},\ and\ \citenamefont
  {Wu}}]{Yang2014}%
  \BibitemOpen
  \bibfield  {author} {\bibinfo {author} {\bibfnamefont {T.-J.}\ \bibnamefont
  {Yang}}, \bibinfo {author} {\bibfnamefont {R.}~\bibnamefont {Shivaraman}},
  \bibinfo {author} {\bibfnamefont {J.~S.}\ \bibnamefont {Speck}}, \ and\
  \bibinfo {author} {\bibfnamefont {Y.-R.}\ \bibnamefont {Wu}},\ }\href
  {\doibase 10.1063/1.4896103} {\bibfield  {journal} {\bibinfo  {journal} {J.
  Appl. Phys.}\ }\textbf {\bibinfo {volume} {116}},\ \bibinfo {pages} {113104}
  (\bibinfo {year} {2014})}\BibitemShut {NoStop}%
\bibitem [{\citenamefont {Schulz}\ \emph {et~al.}(2015)\citenamefont {Schulz},
  \citenamefont {Caro}, \citenamefont {Coughlan},\ and\ \citenamefont
  {O'Reilly}}]{Schulz2015}%
  \BibitemOpen
  \bibfield  {author} {\bibinfo {author} {\bibfnamefont {S.}~\bibnamefont
  {Schulz}}, \bibinfo {author} {\bibfnamefont {M.~A.}\ \bibnamefont {Caro}},
  \bibinfo {author} {\bibfnamefont {C.}~\bibnamefont {Coughlan}}, \ and\
  \bibinfo {author} {\bibfnamefont {E.~P.}\ \bibnamefont {O'Reilly}},\ }\href
  {\doibase 10.1103/PhysRevB.91.035439} {\bibfield  {journal} {\bibinfo
  {journal} {Phys. Rev. B}\ }\textbf {\bibinfo {volume} {91}},\ \bibinfo
  {pages} {035439} (\bibinfo {year} {2015})}\BibitemShut {NoStop}%
\bibitem [{\citenamefont {Maur}\ \emph {et~al.}(2016)\citenamefont {Maur},
  \citenamefont {Pecchia}, \citenamefont {Penazzi}, \citenamefont {Rodrigues},\
  and\ \citenamefont {Di~Carlo}}]{Maur2016}%
  \BibitemOpen
  \bibfield  {author} {\bibinfo {author} {\bibfnamefont {M.~A.~D.}\
  \bibnamefont {Maur}}, \bibinfo {author} {\bibfnamefont {A.}~\bibnamefont
  {Pecchia}}, \bibinfo {author} {\bibfnamefont {G.}~\bibnamefont {Penazzi}},
  \bibinfo {author} {\bibfnamefont {W.}~\bibnamefont {Rodrigues}}, \ and\
  \bibinfo {author} {\bibfnamefont {A.}~\bibnamefont {Di~Carlo}},\ }\href
  {\doibase 10.1103/PhysRevLett.116.027401} {\bibfield  {journal} {\bibinfo
  {journal} {Phys. Rev. Lett.}\ }\textbf {\bibinfo {volume} {116}},\ \bibinfo
  {pages} {027401} (\bibinfo {year} {2016})}\BibitemShut {NoStop}%
\bibitem [{\citenamefont {Filoche}\ \emph {et~al.}(2017)\citenamefont
  {Filoche}, \citenamefont {Piccardo}, \citenamefont {Wu}, \citenamefont {Li},
  \citenamefont {Weisbuch},\ and\ \citenamefont {Mayboroda}}]{Filoche2017}%
  \BibitemOpen
  \bibfield  {author} {\bibinfo {author} {\bibfnamefont {M.}~\bibnamefont
  {Filoche}}, \bibinfo {author} {\bibfnamefont {M.}~\bibnamefont {Piccardo}},
  \bibinfo {author} {\bibfnamefont {Y.-R.}\ \bibnamefont {Wu}}, \bibinfo
  {author} {\bibfnamefont {C.-K.}\ \bibnamefont {Li}}, \bibinfo {author}
  {\bibfnamefont {C.}~\bibnamefont {Weisbuch}}, \ and\ \bibinfo {author}
  {\bibfnamefont {S.}~\bibnamefont {Mayboroda}},\ }\href {\doibase
  10.1103/PhysRevB.95.144204} {\bibfield  {journal} {\bibinfo  {journal} {Phys.
  Rev. B}\ }\textbf {\bibinfo {volume} {95}},\ \bibinfo {pages} {144204}
  (\bibinfo {year} {2017})}\BibitemShut {NoStop}%
\bibitem [{\citenamefont {Piccardo}\ \emph {et~al.}(2017)\citenamefont
  {Piccardo}, \citenamefont {Li}, \citenamefont {Wu}, \citenamefont {Speck},
  \citenamefont {Bonef}, \citenamefont {Farrell}, \citenamefont {Filoche},
  \citenamefont {Martinelli}, \citenamefont {Peretti},\ and\ \citenamefont
  {Weisbuch}}]{Piccardo2017}%
  \BibitemOpen
  \bibfield  {author} {\bibinfo {author} {\bibfnamefont {M.}~\bibnamefont
  {Piccardo}}, \bibinfo {author} {\bibfnamefont {C.-K.}\ \bibnamefont {Li}},
  \bibinfo {author} {\bibfnamefont {Y.-R.}\ \bibnamefont {Wu}}, \bibinfo
  {author} {\bibfnamefont {J.~S.}\ \bibnamefont {Speck}}, \bibinfo {author}
  {\bibfnamefont {B.}~\bibnamefont {Bonef}}, \bibinfo {author} {\bibfnamefont
  {R.~M.}\ \bibnamefont {Farrell}}, \bibinfo {author} {\bibfnamefont
  {M.}~\bibnamefont {Filoche}}, \bibinfo {author} {\bibfnamefont
  {L.}~\bibnamefont {Martinelli}}, \bibinfo {author} {\bibfnamefont
  {J.}~\bibnamefont {Peretti}}, \ and\ \bibinfo {author} {\bibfnamefont
  {C.}~\bibnamefont {Weisbuch}},\ }\href {\doibase 10.1103/PhysRevB.95.144205}
  {\bibfield  {journal} {\bibinfo  {journal} {Phys. Rev. B}\ }\textbf {\bibinfo
  {volume} {95}},\ \bibinfo {pages} {144205} (\bibinfo {year}
  {2017})}\BibitemShut {NoStop}%
\bibitem [{\citenamefont {Li}\ \emph {et~al.}(2017)\citenamefont {Li},
  \citenamefont {Piccardo}, \citenamefont {Lu}, \citenamefont {Mayboroda},
  \citenamefont {Martinelli}, \citenamefont {Peretti}, \citenamefont {Speck},
  \citenamefont {Weisbuch}, \citenamefont {Filoche},\ and\ \citenamefont
  {Wu}}]{Li2017}%
  \BibitemOpen
  \bibfield  {author} {\bibinfo {author} {\bibfnamefont {C.-K.}\ \bibnamefont
  {Li}}, \bibinfo {author} {\bibfnamefont {M.}~\bibnamefont {Piccardo}},
  \bibinfo {author} {\bibfnamefont {L.-S.}\ \bibnamefont {Lu}}, \bibinfo
  {author} {\bibfnamefont {S.}~\bibnamefont {Mayboroda}}, \bibinfo {author}
  {\bibfnamefont {L.}~\bibnamefont {Martinelli}}, \bibinfo {author}
  {\bibfnamefont {J.}~\bibnamefont {Peretti}}, \bibinfo {author} {\bibfnamefont
  {J.~S.}\ \bibnamefont {Speck}}, \bibinfo {author} {\bibfnamefont
  {C.}~\bibnamefont {Weisbuch}}, \bibinfo {author} {\bibfnamefont
  {M.}~\bibnamefont {Filoche}}, \ and\ \bibinfo {author} {\bibfnamefont
  {Y.-R.}\ \bibnamefont {Wu}},\ }\href {\doibase 10.1103/PhysRevB.95.144206}
  {\bibfield  {journal} {\bibinfo  {journal} {Phys. Rev. B}\ }\textbf {\bibinfo
  {volume} {95}},\ \bibinfo {pages} {144206} (\bibinfo {year}
  {2017})}\BibitemShut {NoStop}%
\bibitem [{\citenamefont {Holmes}\ \emph {et~al.}(2014)\citenamefont {Holmes},
  \citenamefont {Choi}, \citenamefont {Kako}, \citenamefont {Arita},\ and\
  \citenamefont {Arakawa}}]{Holmes2014}%
  \BibitemOpen
  \bibfield  {author} {\bibinfo {author} {\bibfnamefont {M.~J.}\ \bibnamefont
  {Holmes}}, \bibinfo {author} {\bibfnamefont {K.}~\bibnamefont {Choi}},
  \bibinfo {author} {\bibfnamefont {S.}~\bibnamefont {Kako}}, \bibinfo {author}
  {\bibfnamefont {M.}~\bibnamefont {Arita}}, \ and\ \bibinfo {author}
  {\bibfnamefont {Y.}~\bibnamefont {Arakawa}},\ }\href {\doibase
  10.1021/nl404400d} {\bibfield  {journal} {\bibinfo  {journal} {Nano Lett.}\
  }\textbf {\bibinfo {volume} {14}},\ \bibinfo {pages} {982} (\bibinfo {year}
  {2014})}\BibitemShut {NoStop}%
\bibitem [{\citenamefont {Samuelson}\ \emph {et~al.}(1992)\citenamefont
  {Samuelson}, \citenamefont {Lindahl}, \citenamefont {Montelius},\ and\
  \citenamefont {Pistol}}]{Samuelson1992}%
  \BibitemOpen
  \bibfield  {author} {\bibinfo {author} {\bibfnamefont {L.}~\bibnamefont
  {Samuelson}}, \bibinfo {author} {\bibfnamefont {J.}~\bibnamefont {Lindahl}},
  \bibinfo {author} {\bibfnamefont {L.}~\bibnamefont {Montelius}}, \ and\
  \bibinfo {author} {\bibfnamefont {M.-E.}\ \bibnamefont {Pistol}},\ }\href
  {\doibase 10.1088/0031-8949/1992/T42/026} {\bibfield  {journal} {\bibinfo
  {journal} {Phys. Scripta}\ }\textbf {\bibinfo {volume} {T42}},\ \bibinfo
  {pages} {149} (\bibinfo {year} {1992})}\BibitemShut {NoStop}%
\bibitem [{\citenamefont {Tereshchenko}\ \emph {et~al.}(2004)\citenamefont
  {Tereshchenko}, \citenamefont {Shaibler}, \citenamefont {Yaroshevich},
  \citenamefont {Shevelev}, \citenamefont {Terekhov}, \citenamefont {Lundin},
  \citenamefont {Zavarin},\ and\ \citenamefont
  {Besyul'kin}}]{Tereshchenko2004}%
  \BibitemOpen
  \bibfield  {author} {\bibinfo {author} {\bibfnamefont {O.~E.}\ \bibnamefont
  {Tereshchenko}}, \bibinfo {author} {\bibfnamefont {G.~E.}\ \bibnamefont
  {Shaibler}}, \bibinfo {author} {\bibfnamefont {A.~S.}\ \bibnamefont
  {Yaroshevich}}, \bibinfo {author} {\bibfnamefont {S.~V.}\ \bibnamefont
  {Shevelev}}, \bibinfo {author} {\bibfnamefont {A.~S.}\ \bibnamefont
  {Terekhov}}, \bibinfo {author} {\bibfnamefont {V.~V.}\ \bibnamefont
  {Lundin}}, \bibinfo {author} {\bibfnamefont {E.~E.}\ \bibnamefont {Zavarin}},
  \ and\ \bibinfo {author} {\bibfnamefont {A.~I.}\ \bibnamefont {Besyul'kin}},\
  }\href {\doibase 10.1134/1.1809437} {\bibfield  {journal} {\bibinfo
  {journal} {Phys. Solid State}\ }\textbf {\bibinfo {volume} {46}},\ \bibinfo
  {pages} {1949} (\bibinfo {year} {2004})}\BibitemShut {NoStop}%
\bibitem [{\citenamefont {Visconti}\ \emph {et~al.}(2000)\citenamefont
  {Visconti}, \citenamefont {Jones}, \citenamefont {Reshchikov}, \citenamefont
  {Cingolani}, \citenamefont {Morkoc},\ and\ \citenamefont
  {Molnar}}]{Visconti2000}%
  \BibitemOpen
  \bibfield  {author} {\bibinfo {author} {\bibfnamefont {P.}~\bibnamefont
  {Visconti}}, \bibinfo {author} {\bibfnamefont {K.~M.}\ \bibnamefont {Jones}},
  \bibinfo {author} {\bibfnamefont {M.~A.}\ \bibnamefont {Reshchikov}},
  \bibinfo {author} {\bibfnamefont {R.}~\bibnamefont {Cingolani}}, \bibinfo
  {author} {\bibfnamefont {H.}~\bibnamefont {Morkoc}}, \ and\ \bibinfo {author}
  {\bibfnamefont {R.~J.}\ \bibnamefont {Molnar}},\ }\href {\doibase
  10.1063/1.1329330} {\bibfield  {journal} {\bibinfo  {journal} {Appl. Phys.
  Lett.}\ }\textbf {\bibinfo {volume} {77}},\ \bibinfo {pages} {3532} (\bibinfo
  {year} {2000})}\BibitemShut {NoStop}%
\bibitem [{\citenamefont {Renaud}\ and\ \citenamefont
  {Alvarado}(1991)}]{Renaud1991}%
  \BibitemOpen
  \bibfield  {author} {\bibinfo {author} {\bibfnamefont {P.}~\bibnamefont
  {Renaud}}\ and\ \bibinfo {author} {\bibfnamefont {S.~F.}\ \bibnamefont
  {Alvarado}},\ }\href {\doibase 10.1103/PhysRevB.44.6340} {\bibfield
  {journal} {\bibinfo  {journal} {Phys. Rev. B}\ }\textbf {\bibinfo {volume}
  {44}},\ \bibinfo {pages} {6340} (\bibinfo {year} {1991})}\BibitemShut
  {NoStop}%
\bibitem [{\citenamefont {Piccardo}\ \emph {et~al.}(2014)\citenamefont
  {Piccardo}, \citenamefont {Martinelli}, \citenamefont {Iveland},
  \citenamefont {Young}, \citenamefont {DenBaars}, \citenamefont {Nakamura},
  \citenamefont {Speck}, \citenamefont {Weisbuch},\ and\ \citenamefont
  {Peretti}}]{Piccardo2014}%
  \BibitemOpen
  \bibfield  {author} {\bibinfo {author} {\bibfnamefont {M.}~\bibnamefont
  {Piccardo}}, \bibinfo {author} {\bibfnamefont {L.}~\bibnamefont
  {Martinelli}}, \bibinfo {author} {\bibfnamefont {J.}~\bibnamefont {Iveland}},
  \bibinfo {author} {\bibfnamefont {N.}~\bibnamefont {Young}}, \bibinfo
  {author} {\bibfnamefont {S.~P.}\ \bibnamefont {DenBaars}}, \bibinfo {author}
  {\bibfnamefont {S.}~\bibnamefont {Nakamura}}, \bibinfo {author}
  {\bibfnamefont {J.~S.}\ \bibnamefont {Speck}}, \bibinfo {author}
  {\bibfnamefont {C.}~\bibnamefont {Weisbuch}}, \ and\ \bibinfo {author}
  {\bibfnamefont {J.}~\bibnamefont {Peretti}},\ }\href {\doibase
  10.1103/PhysRevB.89.235124} {\bibfield  {journal} {\bibinfo  {journal} {Phys.
  Rev. B}\ }\textbf {\bibinfo {volume} {89}},\ \bibinfo {pages} {235124}
  (\bibinfo {year} {2014})}\BibitemShut {NoStop}%
\bibitem [{\citenamefont {Marcinkevicius}\ \emph {et~al.}(2016)\citenamefont
  {Marcinkevicius}, \citenamefont {Uzdavinys}, \citenamefont {Foronda},
  \citenamefont {Cohen}, \citenamefont {Weisbuch},\ and\ \citenamefont
  {Speck}}]{Marcinkevicius2016}%
  \BibitemOpen
  \bibfield  {author} {\bibinfo {author} {\bibfnamefont {S.}~\bibnamefont
  {Marcinkevicius}}, \bibinfo {author} {\bibfnamefont {T.~K.}\ \bibnamefont
  {Uzdavinys}}, \bibinfo {author} {\bibfnamefont {H.~M.}\ \bibnamefont
  {Foronda}}, \bibinfo {author} {\bibfnamefont {D.~A.}\ \bibnamefont {Cohen}},
  \bibinfo {author} {\bibfnamefont {C.}~\bibnamefont {Weisbuch}}, \ and\
  \bibinfo {author} {\bibfnamefont {J.~S.}\ \bibnamefont {Speck}},\ }\href
  {\doibase 10.1103/PhysRevB.94.235205} {\bibfield  {journal} {\bibinfo
  {journal} {Phys. Rev. B}\ }\textbf {\bibinfo {volume} {94}},\ \bibinfo
  {pages} {235205} (\bibinfo {year} {2016})}\BibitemShut {NoStop}%
\bibitem [{\citenamefont {Prietsch}(1995)}]{Prietsch1995}%
  \BibitemOpen
  \bibfield  {author} {\bibinfo {author} {\bibfnamefont {M.}~\bibnamefont
  {Prietsch}},\ }\href@noop {} {\bibfield  {journal} {\bibinfo  {journal}
  {Phys. Rep.}\ }\textbf {\bibinfo {volume} {253}},\ \bibinfo {pages} {164}
  (\bibinfo {year} {1995})}\BibitemShut {NoStop}%
\bibitem [{\citenamefont {Oberli}\ \emph {et~al.}(1990)\citenamefont {Oberli},
  \citenamefont {Shah}, \citenamefont {Damen}, \citenamefont {Kuo},
  \citenamefont {Henry}, \citenamefont {Lary},\ and\ \citenamefont
  {Goodnick}}]{Oberli1990}%
  \BibitemOpen
  \bibfield  {author} {\bibinfo {author} {\bibfnamefont {D.}~\bibnamefont
  {Oberli}}, \bibinfo {author} {\bibfnamefont {J.}~\bibnamefont {Shah}},
  \bibinfo {author} {\bibfnamefont {T.}~\bibnamefont {Damen}}, \bibinfo
  {author} {\bibfnamefont {J.}~\bibnamefont {Kuo}}, \bibinfo {author}
  {\bibfnamefont {J.}~\bibnamefont {Henry}}, \bibinfo {author} {\bibfnamefont
  {J.}~\bibnamefont {Lary}}, \ and\ \bibinfo {author} {\bibfnamefont
  {S.}~\bibnamefont {Goodnick}},\ }\href {\doibase 10.1063/1.102525} {\bibfield
   {journal} {\bibinfo  {journal} {Appl. Phys. Lett.}\ }\textbf {\bibinfo
  {volume} {56}},\ \bibinfo {pages} {1239} (\bibinfo {year}
  {1990})}\BibitemShut {NoStop}%
\bibitem [{\citenamefont {Tada}\ \emph {et~al.}(1988)\citenamefont {Tada},
  \citenamefont {Yamaguchi}, \citenamefont {Ninomiya}, \citenamefont {Uchiki},
  \citenamefont {Kobayashi},\ and\ \citenamefont {YAO}}]{Tada1988}%
  \BibitemOpen
  \bibfield  {author} {\bibinfo {author} {\bibfnamefont {T.}~\bibnamefont
  {Tada}}, \bibinfo {author} {\bibfnamefont {A.}~\bibnamefont {Yamaguchi}},
  \bibinfo {author} {\bibfnamefont {T.}~\bibnamefont {Ninomiya}}, \bibinfo
  {author} {\bibfnamefont {H.}~\bibnamefont {Uchiki}}, \bibinfo {author}
  {\bibfnamefont {T.}~\bibnamefont {Kobayashi}}, \ and\ \bibinfo {author}
  {\bibfnamefont {T.}~\bibnamefont {YAO}},\ }\href {\doibase 10.1063/1.340374}
  {\bibfield  {journal} {\bibinfo  {journal} {J. Appl. Phys.}\ }\textbf
  {\bibinfo {volume} {63}},\ \bibinfo {pages} {5491} (\bibinfo {year}
  {1988})}\BibitemShut {NoStop}%
\bibitem [{\citenamefont {Casey}\ and\ \citenamefont
  {Stern}(1976)}]{Casey1976}%
  \BibitemOpen
  \bibfield  {author} {\bibinfo {author} {\bibfnamefont {H.}~\bibnamefont
  {Casey}}\ and\ \bibinfo {author} {\bibfnamefont {F.}~\bibnamefont {Stern}},\
  }\href {\doibase 10.1063/1.322626} {\bibfield  {journal} {\bibinfo  {journal}
  {J. Appl. Phys.}\ }\textbf {\bibinfo {volume} {47}},\ \bibinfo {pages} {631}
  (\bibinfo {year} {1976})}\BibitemShut {NoStop}%
\bibitem [{\citenamefont {David}\ and\ \citenamefont
  {Grundmann}(2010)}]{David2010}%
  \BibitemOpen
  \bibfield  {author} {\bibinfo {author} {\bibfnamefont {A.}~\bibnamefont
  {David}}\ and\ \bibinfo {author} {\bibfnamefont {M.~J.}\ \bibnamefont
  {Grundmann}},\ }\href {\doibase 10.1063/1.3330870} {\bibfield  {journal}
  {\bibinfo  {journal} {Appl. Phys. Lett.}\ }\textbf {\bibinfo {volume} {96}},\
  \bibinfo {pages} {103504} (\bibinfo {year} {2010})}\BibitemShut {NoStop}%
\bibitem [{\citenamefont {Danhof}\ \emph {et~al.}(2011)\citenamefont {Danhof},
  \citenamefont {Schwarz}, \citenamefont {Kaneta},\ and\ \citenamefont
  {Kawakami}}]{Danhof2011}%
  \BibitemOpen
  \bibfield  {author} {\bibinfo {author} {\bibfnamefont {J.}~\bibnamefont
  {Danhof}}, \bibinfo {author} {\bibfnamefont {U.~T.}\ \bibnamefont {Schwarz}},
  \bibinfo {author} {\bibfnamefont {A.}~\bibnamefont {Kaneta}}, \ and\ \bibinfo
  {author} {\bibfnamefont {Y.}~\bibnamefont {Kawakami}},\ }\href {\doibase
  10.1103/PhysRevB.84.035324} {\bibfield  {journal} {\bibinfo  {journal}
  {Physical Review B}\ }\textbf {\bibinfo {volume} {84}},\ \bibinfo {pages}
  {035324} (\bibinfo {year} {2011})}\BibitemShut {NoStop}%
\bibitem [{\citenamefont {Mensi}\ \emph {et~al.}(2018)\citenamefont {Mensi},
  \citenamefont {Ivanov}, \citenamefont {Uzdavinys}, \citenamefont {Kelchner},
  \citenamefont {Nakamura}, \citenamefont {DenBaars}, \citenamefont {Speck},\
  and\ \citenamefont {Marcinkevicius}}]{Mensi2018}%
  \BibitemOpen
  \bibfield  {author} {\bibinfo {author} {\bibfnamefont {M.}~\bibnamefont
  {Mensi}}, \bibinfo {author} {\bibfnamefont {R.}~\bibnamefont {Ivanov}},
  \bibinfo {author} {\bibfnamefont {T.~K.}\ \bibnamefont {Uzdavinys}}, \bibinfo
  {author} {\bibfnamefont {K.~M.}\ \bibnamefont {Kelchner}}, \bibinfo {author}
  {\bibfnamefont {S.}~\bibnamefont {Nakamura}}, \bibinfo {author}
  {\bibfnamefont {S.~P.}\ \bibnamefont {DenBaars}}, \bibinfo {author}
  {\bibfnamefont {J.~S.}\ \bibnamefont {Speck}}, \ and\ \bibinfo {author}
  {\bibfnamefont {S.}~\bibnamefont {Marcinkevicius}},\ }\href {\doibase
  10.1021/acsphotonics.7b01061} {\bibfield  {journal} {\bibinfo  {journal} {Acs
  Photonics}\ }\textbf {\bibinfo {volume} {5}},\ \bibinfo {pages} {528}
  (\bibinfo {year} {2018})}\BibitemShut {NoStop}%
\bibitem [{\citenamefont {Hegarty}\ and\ \citenamefont
  {Sturge}(1985)}]{Hegarty1985}%
  \BibitemOpen
  \bibfield  {author} {\bibinfo {author} {\bibfnamefont {J.}~\bibnamefont
  {Hegarty}}\ and\ \bibinfo {author} {\bibfnamefont {M.~D.}\ \bibnamefont
  {Sturge}},\ }\href@noop {} {\bibfield  {journal} {\bibinfo  {journal} {J.
  Opt. Soc. Am. B}\ }\textbf {\bibinfo {volume} {2}},\ \bibinfo {pages} {1143}
  (\bibinfo {year} {1985})}\BibitemShut {NoStop}%
\bibitem [{\citenamefont {Kaiser}\ \emph {et~al.}(2001)\citenamefont {Kaiser},
  \citenamefont {Bell}, \citenamefont {Hecht},\ and\ \citenamefont
  {Davis}}]{Kaiser2001}%
  \BibitemOpen
  \bibfield  {author} {\bibinfo {author} {\bibfnamefont {W.~J.}\ \bibnamefont
  {Kaiser}}, \bibinfo {author} {\bibfnamefont {L.~D.}\ \bibnamefont {Bell}},
  \bibinfo {author} {\bibfnamefont {M.~H.}\ \bibnamefont {Hecht}}, \ and\
  \bibinfo {author} {\bibfnamefont {L.~C.}\ \bibnamefont {Davis}},\ }\href@noop
  {} {\emph {\bibinfo {title} {Scanning Probe Microscopy and Spectroscopy:
  Theory, Techniques, and Applications}}},\ \bibinfo {edition} {2nd}\ ed.,\
  edited by\ \bibinfo {editor} {\bibfnamefont {D.~A.}\ \bibnamefont {Bonnell}}\
  (\bibinfo  {publisher} {Wiley/VCH New York},\ \bibinfo {year} {2001})\ p.\
  \bibinfo {pages} {253}\BibitemShut {NoStop}%
\bibitem [{\citenamefont {Evoy}\ \emph {et~al.}(1999)\citenamefont {Evoy},
  \citenamefont {Harnett}, \citenamefont {Craighead}, \citenamefont {Keller},
  \citenamefont {Mishra},\ and\ \citenamefont {DenBaars}}]{Evoy1999}%
  \BibitemOpen
  \bibfield  {author} {\bibinfo {author} {\bibfnamefont {S.}~\bibnamefont
  {Evoy}}, \bibinfo {author} {\bibfnamefont {C.~K.}\ \bibnamefont {Harnett}},
  \bibinfo {author} {\bibfnamefont {H.~G.}\ \bibnamefont {Craighead}}, \bibinfo
  {author} {\bibfnamefont {S.}~\bibnamefont {Keller}}, \bibinfo {author}
  {\bibfnamefont {U.~K.}\ \bibnamefont {Mishra}}, \ and\ \bibinfo {author}
  {\bibfnamefont {S.~P.}\ \bibnamefont {DenBaars}},\ }\href {\doibase
  10.1063/1.123580} {\bibfield  {journal} {\bibinfo  {journal} {Appl. Phys.
  Lett.}\ }\textbf {\bibinfo {volume} {74}},\ \bibinfo {pages} {1457} (\bibinfo
  {year} {1999})}\BibitemShut {NoStop}%
\bibitem [{\citenamefont {Manson-Smith}\ \emph {et~al.}(2001)\citenamefont
  {Manson-Smith}, \citenamefont {Trager-Cowan},\ and\ \citenamefont
  {O'Donnell}}]{Manson-Smith2001}%
  \BibitemOpen
  \bibfield  {author} {\bibinfo {author} {\bibfnamefont {S.}~\bibnamefont
  {Manson-Smith}}, \bibinfo {author} {\bibfnamefont {C.}~\bibnamefont
  {Trager-Cowan}}, \ and\ \bibinfo {author} {\bibfnamefont {K.}~\bibnamefont
  {O'Donnell}},\ }\href {\doibase
  10.1002/1521-3951(200111)228:2<445::AID-PSSB445>3.0.CO;2-I} {\bibfield
  {journal} {\bibinfo  {journal} {Phys. Status Solidi B}\ }\textbf {\bibinfo
  {volume} {228}},\ \bibinfo {pages} {445} (\bibinfo {year}
  {2001})}\BibitemShut {NoStop}%
\bibitem [{\citenamefont {Filoche}\ and\ \citenamefont
  {Mayboroda}(2012)}]{Filoche2012}%
  \BibitemOpen
  \bibfield  {author} {\bibinfo {author} {\bibfnamefont {M.}~\bibnamefont
  {Filoche}}\ and\ \bibinfo {author} {\bibfnamefont {S.}~\bibnamefont
  {Mayboroda}},\ }\href {\doibase 10.1073/pnas.1120432109} {\bibfield
  {journal} {\bibinfo  {journal} {P. Natl. Acad. Sci. USA}\ }\textbf {\bibinfo
  {volume} {109}},\ \bibinfo {pages} {14761} (\bibinfo {year}
  {2012})}\BibitemShut {NoStop}%
\bibitem [{\citenamefont {Arnold}\ \emph {et~al.}(2016)\citenamefont {Arnold},
  \citenamefont {David}, \citenamefont {Jerison}, \citenamefont {Mayboroda},\
  and\ \citenamefont {Filoche}}]{Arnold2016}%
  \BibitemOpen
  \bibfield  {author} {\bibinfo {author} {\bibfnamefont {D.~N.}\ \bibnamefont
  {Arnold}}, \bibinfo {author} {\bibfnamefont {G.}~\bibnamefont {David}},
  \bibinfo {author} {\bibfnamefont {D.}~\bibnamefont {Jerison}}, \bibinfo
  {author} {\bibfnamefont {S.}~\bibnamefont {Mayboroda}}, \ and\ \bibinfo
  {author} {\bibfnamefont {M.}~\bibnamefont {Filoche}},\ }\href {\doibase
  10.1103/PhysRevLett.116.056602} {\bibfield  {journal} {\bibinfo  {journal}
  {Phys. Rev. Lett.}\ }\textbf {\bibinfo {volume} {116}},\ \bibinfo {pages}
  {056602} (\bibinfo {year} {2016})}\BibitemShut {NoStop}%
\end{thebibliography}%


\begin{thebibliography}{2}%
\makeatletter
\providecommand \@ifxundefined [1]{%
 \@ifx{#1\undefined}
}%
\providecommand \@ifnum [1]{%
 \ifnum #1\expandafter \@firstoftwo
 \else \expandafter \@secondoftwo
 \fi
}%
\providecommand \@ifx [1]{%
 \ifx #1\expandafter \@firstoftwo
 \else \expandafter \@secondoftwo
 \fi
}%
\providecommand \natexlab [1]{#1}%
\providecommand \enquote  [1]{``#1''}%
\providecommand \bibnamefont  [1]{#1}%
\providecommand \bibfnamefont [1]{#1}%
\providecommand \citenamefont [1]{#1}%
\providecommand \href@noop [0]{\@secondoftwo}%
\providecommand \href [0]{\begingroup \@sanitize@url \@href}%
\providecommand \@href[1]{\@@startlink{#1}\@@href}%
\providecommand \@@href[1]{\endgroup#1\@@endlink}%
\providecommand \@sanitize@url [0]{\catcode `\\12\catcode `\$12\catcode
  `\&12\catcode `\#12\catcode `\^12\catcode `\_12\catcode `\%12\relax}%
\providecommand \@@startlink[1]{}%
\providecommand \@@endlink[0]{}%
\providecommand \url  [0]{\begingroup\@sanitize@url \@url }%
\providecommand \@url [1]{\endgroup\@href {#1}{\urlprefix }}%
\providecommand \urlprefix  [0]{URL }%
\providecommand \Eprint [0]{\href }%
\providecommand \doibase [0]{http://dx.doi.org/}%
\providecommand \selectlanguage [0]{\@gobble}%
\providecommand \bibinfo  [0]{\@secondoftwo}%
\providecommand \bibfield  [0]{\@secondoftwo}%
\providecommand \translation [1]{[#1]}%
\providecommand \BibitemOpen [0]{}%
\providecommand \bibitemStop [0]{}%
\providecommand \bibitemNoStop [0]{.\EOS\space}%
\providecommand \EOS [0]{\spacefactor3000\relax}%
\providecommand \BibitemShut  [1]{\csname bibitem#1\endcsname}%
\let\auto@bib@innerbib\@empty
\bibitem [{\citenamefont {Tereshchenko}\ \emph {et~al.}(2004)\citenamefont
  {Tereshchenko}, \citenamefont {Shaibler}, \citenamefont {Yaroshevich},
  \citenamefont {Shevelev}, \citenamefont {Terekhov}, \citenamefont {Lundin},
  \citenamefont {Zavarin},\ and\ \citenamefont
  {Besyul'kin}}]{Tereshchenko2004}%
  \BibitemOpen
  \bibfield  {author} {\bibinfo {author} {\bibfnamefont {O.~E.}\ \bibnamefont
  {Tereshchenko}}, \bibinfo {author} {\bibfnamefont {G.~E.}\ \bibnamefont
  {Shaibler}}, \bibinfo {author} {\bibfnamefont {A.~S.}\ \bibnamefont
  {Yaroshevich}}, \bibinfo {author} {\bibfnamefont {S.~V.}\ \bibnamefont
  {Shevelev}}, \bibinfo {author} {\bibfnamefont {A.~S.}\ \bibnamefont
  {Terekhov}}, \bibinfo {author} {\bibfnamefont {V.~V.}\ \bibnamefont
  {Lundin}}, \bibinfo {author} {\bibfnamefont {E.~E.}\ \bibnamefont {Zavarin}},
  \ and\ \bibinfo {author} {\bibfnamefont {A.~I.}\ \bibnamefont {Besyul'kin}},\
  }\href {\doibase 10.1134/1.1809437} {\bibfield  {journal} {\bibinfo
  {journal} {Phys. Solid State}\ }\textbf {\bibinfo {volume} {46}},\ \bibinfo
  {pages} {1949} (\bibinfo {year} {2004})}\BibitemShut {NoStop}%
\bibitem [{\citenamefont {Visconti}\ \emph {et~al.}(2000)\citenamefont
  {Visconti}, \citenamefont {Jones}, \citenamefont {Reshchikov}, \citenamefont
  {Cingolani}, \citenamefont {Morkoc},\ and\ \citenamefont
  {Molnar}}]{Visconti2000}%
  \BibitemOpen
  \bibfield  {author} {\bibinfo {author} {\bibfnamefont {P.}~\bibnamefont
  {Visconti}}, \bibinfo {author} {\bibfnamefont {K.~M.}\ \bibnamefont {Jones}},
  \bibinfo {author} {\bibfnamefont {M.~A.}\ \bibnamefont {Reshchikov}},
  \bibinfo {author} {\bibfnamefont {R.}~\bibnamefont {Cingolani}}, \bibinfo
  {author} {\bibfnamefont {H.}~\bibnamefont {Morkoc}}, \ and\ \bibinfo {author}
  {\bibfnamefont {R.~J.}\ \bibnamefont {Molnar}},\ }\href {\doibase
  10.1063/1.1329330} {\bibfield  {journal} {\bibinfo  {journal} {Appl. Phys.
  Lett.}\ }\textbf {\bibinfo {volume} {77}},\ \bibinfo {pages} {3532} (\bibinfo
  {year} {2000})}\BibitemShut {NoStop}%
\end{thebibliography}%

\end{document}